\documentclass[a4paper,reqno]{amsart}
\usepackage{amssymb}
\usepackage{latexsym}
\usepackage{amsmath}
\usepackage{euscript}
\usepackage{graphics,color}
\usepackage[all]{xy}
\usepackage[margin=3cm]{geometry}
\usepackage{MnSymbol} % for \udots

   \def\dZ{{\mathbb Z}}

\def\cA{{\mathcal A}}      
      
      \def\cI{{\mathcal I}}

      \def\cR{{\mathcal R}}
   \def\cT{{\mathcal T}}   
   \def\cW{{\mathcal W}}

\newcommand{\be}{\begin{equation}}
\newcommand{\ee}{\end{equation}}
\newcommand{\ba}{\begin{eqnarray}}
\newcommand{\ea}{\end{eqnarray}}
\newcommand{\baa}{\begin{eqnarray*}}
\newcommand{\eaa}{\end{eqnarray*}}
\newcommand{\bb}{\begin{thebibliography}}
\newcommand{\eb}{\end{thebibliography}}
\newcommand{\lab}[1]{\label{#1}}
\newcommand{\re}[1]{(\ref{#1})}

\newcounter{my}
\newcommand{\he}%
   {\stepcounter{equation}\setcounter{my}%
   {\value{equation}}\setcounter{equation}0%
   }%
\newcommand{\she}%
   {\setcounter{equation}{\value{my}}%
    }%

\renewcommand\t{\tilde}

\newtheorem{theorem}{Theorem}[section]

\newtheorem{definition}[theorem]{Definition}
\theoremstyle{definition}

\newtheorem{remark}[theorem]{Remark}

\numberwithin{equation}{section}

\begin{document}

\title{Ruijsenaars-van Diejen-Takemura Hamiltonians as rational Heun operators}

\author{Satoshi Tsujimoto}
\author{Luc Vinet}
\author{Alexei Zhedanov}

\address{Graduate School of Informatics, Kyoto University,
Yoshida-Honmachi, Kyoto, Japan 606-8501}

\address{Centre de recherches \\ math\'ematiques and IVADO,
Universit\'e de Montr\'eal, P.O. Box 6128, Centre-ville Station,
Montr\'eal (Qu\'ebec), H3C 3J7}

\address{Leonhard Euler International Mathematical Institute, Saint Petersburg, Russian Federation}

\begin{abstract}
The most general Ruijsenaars-van Diejen-Takemura Hamiltonians are characterized as Heun operators defined as second order $q$-difference operators with a raising action on elementary rational functions with poles on the Askey-Wilson grid.
\end{abstract}


\maketitle

\section{Introduction}
We consider the generalized Heun operators that are defined as raising operators on sets of elementary rational functions with poles on Askey-Wilson grids. It will be found that these operators coincide with Hamiltonians already identified in the context of certain integrable models. We shall show that this connection can be established by considering rational Heun operators associated to either one \cite{tsujimoto2023rational}, or equivalently two, series of poles. Here is some background to start.

The elliptic integrable quantum many-body $BC_N$-models of Ruijsenaars \cite{ruijsenaars2004integrable} and van Diejen \cite{van1994integrability} are relativistic extensions of the Inozemtsev systems \cite{inozemtsev1989lax}. Their Hamiltonians are given in terms of $q$-difference operators. Takemura has looked at their one-particle ($N=1$) specialization and has given \cite{takemura2017degenerations} four degenerations  $\mathcal{A}^{(i)}, \;  i=1,2,3,4,$ of the one-body Hamiltonian that are interesting on their own. We shall refer to these as the RvDT Hamiltonians.  The expression (in our notations) of the most general one, $\mathcal{A}^{(1)}$, will be given in the body of the article. We moreover indicate in Appendix A how this operator is obtained from its elliptic version given in \cite{atai_noumi2023}.

The Heun equation \cite{kristensson2010second} is the standardized second order Fuchsian differential equation with four regular singularities. It is known \cite{oshima1995commuting} \cite{takemura2003heun} that finding the eigenstates of the Inozemtsev $BC_1$-Hamiltonian amounts to solving the Heun equation. This can be done using the Bethe ansatz approach \cite{takemura2003heun}. (See also \cite{bernard2021heun}.) Therefore, the Schr\"odinger equation of the Inozemtsev model can be considered as a multivariate extension of the standard Heun equation. This line of thought leads one to view the Ruijsenaars-van Diejen models as providing multivariate $q$-extensions of the Heun equation and hence, Takemura's degenerations of the one-particle relativistic Hamiltonian stand as various $q$-versions of the univariate differential Heun equation.

The Heun equation has connections with diverse areas of mathematics and in particular with the Painlev\'e equations through the ``antiquantization'' procedure \cite{slavyanov2015antiquantization}.
This relation has been exploited to explore $q$-Heun equations \cite{takemura2003heun}. Recently in collaboration with others, the authors have developed an approach to the identification and study of (second order differential or difference) operators of Heun type that is based on certain raising properties of these operators on sets of elementary functions. The set of monomials was considered first to obtain Heun operators associated to the polynomials of the Askey scheme \cite{koekoek2010hypergeometric}. The scope was enlarged subsequently \cite{tsujimoto2023rational} by examining actions on sets of basic rational functions. 

The most general second order $q$-difference operator that raises by one the degree of polynomials defined on the exponential lattice was thus constructed and is referred to as the $q$-Heun operator of big $q$-Jacobi type \cite{baseilhac2020q}. It was observed that this $q$-Heun operator coincides with the RvDT Hamiltonian $\mathcal{A}^{(3)}$. Upon specializing to the little 
$q$-Jacobi case, one obtains the $q$-Heun operator that Hahn had introduced \cite{hahn1971linear} and that coincides with the RvDT Hamiltonian $\mathcal{A}^{(4)}$. While this supported the idea that the RvDT Hamiltonians are certain Heun operators, it left open the similar identification of the least degenerate Hamiltonians $\mathcal{A}^{(1)}$ and $\mathcal{A}^{(2)}$ as  Heun operators.

This motivated in part the exploration of Heun operators defined through a raising property on rational functions. 
A program was hence undertaken in \cite{tsujimoto2023rational} where the rational Heun operator were defined as the most general second order $q$-difference operator which sends any rational function of type $[(n-1)/n]$ to a rational function of type $[n/(n+1)]$. These operators were constructed,
by looking at their actions on elementary rational functions $(x-x_n)^{-1}, n=0,1,\ldots$. with poles on the Askey-Wilson grid $x_n = \alpha q^n + (\alpha q^n)^{-1}$. However, the connection with the RvDT Hamiltonians $\mathcal{A}^{(1)}$ was not  fully established and this is what is accomplished here in two ways. On the one hand, we shall indicate how $\mathcal{A}^{(1)}$ is obtained from this raising Heun operator associated to one series of poles through a parameter inversion and a gauge transformation. On the other hand, following an idea raised in \cite{tsujimoto2023rational}, we shall find that $\mathcal{A}^{(1)}$ can also be recovered from the raising Heun operator associated to two series of poles. The identification of $\mathcal{A}^{(2)}$ as a rational Heun operator in this framework requires the poles to lie on a $q$-linear grid and will be treated in detail separately \cite{TVZ_qlin}.

We may remark that in the case of Heun operators associated to polynomial families of the Askey scheme, the definition of these operators as raising by one the degree of monomials defined on the corresponding lattices coincides with the generic bilinear expression in the bispectral operators of the polynomials (the associated algebraic Heun operator\cite{grunbaum2018algebraic} or the tridiagonalization process). Such a dual approach is not known at the moment for rational Heun operators but could be rooted in the Leonard trio theory recently developed in \cite{CGLMVW_Trios}. We may further add that the Heun-Askey-Wilson operator which arises from a raising condition on polynomials on the Askey-Wilson grid was shown in \cite{tsujimoto2023rational} to be obtainable as a limit of $\mathcal{A}^{(1)}$.

With the goal of constructing Heun operators defined as second order $q$-difference operators that have a raising action on one or two sets of elementary rational functions whose poles are on Askey-Wilson grids, the plan of the paper is the following. In section 2 we make explicit the two defining raising conditions. The characterization of the associated Heun operators will proceed, given the generic form for these operators, by enforcing necessary conditions stemming from the prescribed raising actions on a minimal set of elementary rational functions. The sufficiency of these conditions will then follow a posteriori by verifying (particularly in the two-pole series) that the operator thus constructed is uniquely defined and has the desired action on generic elementary rational functions. This program will be implemented by first determining in Section 3 an operator $W_{AW}$ that subsumes the two cases of interest. The desired Heun operators will then be constructed by further specializations in the next two sections: the operator $W^{(1)}_{AW}$ associated to one series of poles in Section 4 and the operator $W^{(2)}_{AW}$ attached to two series of poles in Section 5. This is where the sufficiency of the necessary conditions that will have been used will be addressed. The identification of $W^{(2)}_{AW}$ with the Takemura Hamiltonian $\mathcal{A}^{(1)}$ will be made in Section 6 where the equivalence between $W^{(1)}_{AW}$ and $W^{(2)}_{AW}$  will also be stated, thus establishing that the Heun operator defined with a single series of poles also happens to essentially be the Takemura operator $\mathcal{A}^{(1)}$.
The paper will end with short concluding remarks and an outlook that are followed by the Appendix A mentioned before.

\section{Raising properties}

\begin{definition}
A rational function $R(x)$ is said to be of \emph{type $[r/s]$} if it can be written as
\be
R(x)=\frac{P(x)}{Q(x)},
\ee
where $P$ and $Q$ are coprime polynomials, and  $\deg P= r,\ \deg Q = s$. \\

\noindent Note: In the following we shall restrict to the simple situation where $Q$ has only simple zeros.
\end{definition}

Take $R_{n,m}(x)$ a rational function of type $[(n+m+1)/(n+m+2)]$ in the variable $x=x(z)$,
whose poles are prescribed at
$x=x_0,x_1,\dots, x_n$ and $x=y_0,y_1,\dots,y_m$.
Since only simple poles are allowed, such a function admits a partial-fraction expansion of the form
\be
R_{n,m}(x)=R_n(x;x_0,x_1,\ldots)+R_m'(x;y_0,y_1,\ldots),
\lab{expansion_Rnm}
\ee
where
\ba
&&R_n(x)=\frac{\xi_0}{x-x_0}+\frac{\xi_1}{x-x_1}+\cdots+\frac{\xi_n}{x-x_n},
\lab{expansion_Rn}\\
&&R_m'(x)=\frac{\xi_0'}{x-y_0}+\frac{\xi_1'}{x-y_1}+\cdots+\frac{\xi_m'}{x-y_m}.
\lab{expansion_Rm}
\ea
We use the conventions $R_{n,-1}(x)=R_n(x)$ and $R_{-1,m}(x)=R_m'(x)$.

Bases of elementary rational functions labeled by the sets of ordered poles $\{x_n\}_{n \in \dZ_{\ge 0}}$ and $\{y_n\}_{n \in \dZ_{\ge 0}}$ read:

\be
\left\{\frac{1}{x-x_0}, \frac{1}{x-x_1}, \ldots\right\}; \qquad \left\{\frac{1}{x-y_0}, \frac{1}{x-y_1}, \ldots\right\}.
\ee

The functions $R_n(x)$, $R_m'(x)$, and $R_{n,m}(x)$ introduced above are thus given as expansions in terms of these basis elements. The raising properties that we will describe next will proceed with respect to the prescribed orderings.

Consider the generic second order $q$-difference operator
\be
W= A_1(z) \cT^{+} + A_2(z) \cT^{-} + A_0(z) \cI ,
\lab{gen_W}
\ee
where $\cT^{+} f(z) = f(zq)$, $\cT^{-} f(z) = f(z/q)$, and $\cI f(z)=f(z)$.

In contrast with the polynomial case, there are several types of  degree-raising property for rational functions.
We will consider the following two cases:
\begin{enumerate}

 \item[(i)] one series of poles $\{x_n\}_{n \in \dZ_{\ge 0}}$: 
\be
\lab{type2:cond}
W^{(1)} R_n(x) = \t R_{n+1}(x) =  \frac{\t\xi_0}{x-x_0}  + \dots +\frac{\t\xi_n}{x-x_n} + \frac{\t\xi_{n+1}}{x-x_{n+1}}, \ee

 \item[(ii)] two series of poles 
 $\{x_n,y_n\}_{n \in \dZ_{\ge 0}}$:

\be
\lab{type3:cond}
W^{(2)} R_{n,m}(x) = \t R_{n+1,m+1}(x) = \frac{\t\xi_0}{x-x_0} + \dots  + \frac{\t\xi_{n+1}}{x-x_{n+1}}+\frac{\t\xi_0'}{x-y_0} + \dots  + \frac{\t\xi'_{m+1}}{x-y_{m+1}},\ee

\end{enumerate}
where the two series of poles are assumed to be independent, that is, $y_n \notin \{x_j\}_{j \in \dZ}$. 

\noindent Note that $W^{(1)}$ adds one pole while $W^{(2)}$ adds two,  and hence generically one is not a special case of the other:
\begin{equation}
    W^{(1)}: R_{n,-1} \rightarrow R_{n+1,-1} \quad (+ 1\; \text{pole}) \qquad \text{and} \qquad  
    W^{(2)}: R_{n,m} \rightarrow R_{n+1,m+1} \quad (+ 2 \;\text{poles})
\end{equation}

\noindent Two rational Heun operators satisfying each of the conditions \re{type2:cond} and \re{type3:cond} will be constructed with the series of poles taken as the case may be on one or two Askey-Wilson grids. With $x(z)=z+1/z$, the series of poles will thus be:
\begin{equation}
    x_n=x(\alpha q^n)=\alpha q^n + \alpha^{-1} q^{-n} \quad \text{and} \quad  y_n=x(\beta q^n)=\beta q^n + \beta^{-1} q^{-n} \quad \text{with}\quad n\in \mathbb{Z}_{\ge 0}, \quad \alpha \neq \beta \in \mathbb{C}.
\end{equation} \label{xnyn}
The corresponding Heun operators will be respectively called $W^{(1)}_{\!\text{AW}}$ and $W^{(2)}_{\!\text{AW}}$.

\section{An intermediary step: the operator $W_{\!\text{AW}}$}
\subsection{General properties}

It will prove convenient as an intermediary step to introduce the operator $W$ defined on functions $R_n \equiv R_{n,-1}$ but acting according the raising property (ii), ie.:
\begin{equation}
     W: R_{n,-1} \rightarrow R_{n+1,0} \quad (+ 2\; \text{poles}).
\end{equation}
In other words, $W$ adds two poles one of which we can call $y_0$. As we shall see, it is simple to realize that $  W^{(1)}$ and $  W^{(2)}$ are here special cases of $W$. We shall therefore begin by characterizing $W$ before obtaing $  W^{(1)}$ and $  W^{(2)}$ by adding appropriate conditions.

By definition the operator $W$ must satisfy the following three conditions:
\ba
 && W\left\{\dfrac{1}{x-x_0}\right\}=\dfrac{\xi_{0,+}}{x-y_0}+\dfrac{\xi_{0,0}}{x-x_0}+\dfrac{\xi_{0,1}}{x-x_1}=r_1(x)\mbox,\label{W:cond1}\\
 && W\left\{\dfrac{1}{x-x_1}\right\}=\dfrac{\xi_{1,+}}{x-y_0}+\dfrac{\xi_{1,0}}{x-x_0}+\dfrac{\xi_{1,1}}{x-x_1}+\dfrac{\xi_{1,2}}{x-x_2}=r_2(x),\label{W:cond2}\\
 && W\left\{\dfrac{1}{x-x_2}\right\}=\dfrac{\xi_{2,+}}{x-y_0}+\dfrac{\xi_{2,0}}{x-x_0}+\dfrac{\xi_{2,1}}{x-x_1}+\dfrac{\xi_{2,2}}{x-x_2}+\dfrac{\xi_{2,3}}{x-x_3}=r_3(x),\label{W:cond3}
\ea
where $x_n$ and $y_n$ are defined as in \eqref{xnyn}.

From (\ref{W:cond1})--(\ref{W:cond3}), we can find the explicit expressions of the functions $A_i(z)$ entering in the expression of $W$ as per \eqref{gen_W}: 
\be A_1(z) = \frac{(x(zq)-x_0)(x(zq)-x_1)(x(zq)-x_2)}{(x(zq)-x(z))(x(zq)-x(z/q))}
\sum_{j=0}^{2}\frac{(x(z)-x_j)(x(z/q)-x_j) }{\prod_{\ell\in\{0,1,2\}\backslash{j}}(x_j-x_{\ell})} r_{j+1}(x(z)), \lab{A1_expl} \ee
\be A_2(z) = \frac{(x(z/q)-x_0)(x(z/q)-x_1)(x(z/q)-x_2)}{(x(z/q)-x(z))(x(z/q)-x(zq))}
\sum_{j=0}^{2}\frac{(x(z)-x_j)(x(zq)-x_j) }{\prod_{\ell\in\{0,1,2\}\backslash{j}}(x_j-x_{\ell})} r_{j+1}(x(z)) \lab{A2_expl},\ee
\be A_0(z) = (x(z)-x_0)r_1(x(z)) - \dfrac{x(z)-x_0}{x(qz)-x_0}A_1(z) -\dfrac{x(z)-x_0}{x(z/q)-x_0}A_2(z) \lab{A0_expl}. \ee
Note that the functions $A_1(z)$  and $A_2(z)$ are related by 
\be A_2(z)=A_1(1/z), \ee
if $x(1/z)=x(z)$ as is the case when $x(z)=z+z^{-1}$.

By taking $\xi_{0,+}=\xi_{1,+}=\xi_{2,+}=0$, the three conditions (\ref{W:cond1})--(\ref{W:cond3}) are those for the raising operator $W^{(1)}$ of type 1 \re{type2:cond}. In order to satisfy the raising operator property of type 2 for all non-negative integers $n,m$, we have to impose additional conditions on the actions to the rational functions with the poles $\{y_n\}_{n\in \dZ_{\ge 0}}$.

\subsection{Restricting $W$ to the Askey-Wilson grid}
Let us now take $W$ on the Askey-Wilson grid with the poles given by \eqref{xnyn}. 
The restriction $W_{\!\text{AW}}$ to that grid of the operator $W$ 
satisfying (\ref{W:cond1})--(\ref{W:cond3}) is found to have the following expression  from \eqref{A1_expl}-\eqref{A0_expl}:
\ba
&& W_{\!\text{AW}} = A_1(z)\cT^{+} +A_2(z)\cT^{-}+A_0(z)\cI,\\
&& A_1(z) = \dfrac{(qz - \alpha) Q_{10}(z)}{z^2 (1-\alpha z )(\beta - z)(1-\beta z)(1-z^2)(1-q z^2)},\\
&& A_0(z) = c_0 +c_1\left(z+\frac{1}{z}\right) + c_2\left(z^2+\frac{1}{z^2}\right) \nonumber\\
&& \phantom{A_0(z) =}+\dfrac{c_3 }{(1-\beta z)(1-\beta/z)}+\frac{c_4}{(1-q^{-\frac12} z)(1-q^{-\frac12}/z)} +\frac{c_5}{(1+q^{-\frac12} z)(1+q^{-\frac12}/z)}
\ea
and $A_2(z)=A_1(1/z)$, where $Q_{10}(z)$ is a polynomial of degree 10 in $z$ defined by
\be
Q_{10}(z) = \sum_{j=0}^{10}\rho_j z^j.
\ee
One can see that all the 17 coefficients $\rho_j$ and $c_j$ are linear combinations of the elements in the set $\Xi = \{\xi_{0,+}, \xi_{1,+}, \xi_{2,+}, \xi_{0,0}$, $\xi_{0,1}$, $\xi_{1,0}$,$ \xi_{1,1}$, $\xi_{1,2}$, $\xi_{2,0}, \xi_{2,1}, \xi_{2,2}, \xi_{2,3}\}$,
\be
\rho_j = \sum_{\xi \in \Xi } d_{j,\xi}(\alpha,\beta,q)\,\xi, \quad 
c_j = \sum_{\xi \in \Xi } \tilde d_{j,\xi}(\alpha,\beta,q)\,\xi,
\ee
and, by examining the relations among them, that the 12 coefficients $\rho_0, \ldots, \rho_9$  and $c_0,c_3$ are linearly independent superpositions with respect to $\Xi$ consisting of 12 parameters while the remaining 5 coefficients can be expressed as superpositions of $\{\rho_0,\rho_1,\ldots,\rho_9\}$ as follows:
\ba
&\rho_{10}=\alpha^2 q^3 \rho_0, \quad 
c_1=q \left( 1+q \right)  \left( 1+\beta^{-2} \right) \alpha \rho_{{0}}+\dfrac {q {\alpha}\rho_{{1}}}{\beta}+\dfrac {\rho_{{9}}
}{q \,\alpha\,\beta}, \quad 
c_2 = \dfrac{q (1+q) \alpha \rho_0}{\beta},\\
& \displaystyle
c_4 = \dfrac{-1}{2q^3(1-\beta q^{-\frac12})(1-\beta q^{\frac12})}\sum_{j=0}^{10} q^{\frac{10-j}{2}}\rho_j, \quad 
 c_5 = \dfrac{-1}{2q^3(1+\beta q^{-\frac12})(1+\beta q^{\frac12})} \sum_{j=0}^{10} (-q)^{\frac{10-j}{2}}\rho_j.
\ea
Now that we have determined the operator $W_{\!\text{AW}}$ , it is straightforward to compute its action on $(x-y_0)^{-1}, (x-y_1)^{-1}$. The results can be presented  in the following form: 
\ba
\lab{W:action_y0}
&& W_{\!\text{AW}}\left\{\dfrac{1}{x-y_0}\right\} = 
\dfrac{\xi_{-1,+}}{x-x_0}+\dfrac{\xi_{-1,-1}}{x-y_{-1}}+\dfrac{\xi_{-1,0}}{x-y_0}+\dfrac{\xi_{-1,0}'}{(x-y_0)^2}+\dfrac{\xi_{-1,1}}{x-y_1},\\
\lab{W:action_y1}
&& W_{\!\text{AW}}\left\{\dfrac{1}{x-y_1}\right\} = 
\dfrac{\xi_{-2,+}}{x-x_0}+\dfrac{\xi_{-2, 0}}{x-y_{ 0}}+\dfrac{\xi_{-2,0}'}{(x-y_0)^2}+\dfrac{\xi_{-2,1}}{x-y_1}+\dfrac{\xi_{-2,2}}{x-y_2}.
\ea

\section{The Heun operator $W^{(1)}_{\!\text{AW}}$ associated to one series of poles on the Askey-Wilson grid}
We now turn to the determination of the rational Heun operator $W^{(1)}_{\!\text{AW}}$ on the Askey-Wilson grid that is defined through the raising property (i) given in 
\re{type2:cond}. Since this operator transforms rational functions of type $[n/(n+1)]$ into those of type $[(n+1)/(n+2)]$,
relative to the action of $W_{\!\text{AW}}$ as given in \eqref{W:cond1} - \eqref{W:cond3}, we need to ensure that the second pole at $y_0$ is not added which is tantamount to imposing the following conditions

 \be\xi_{0,+}=\xi_{1,+}=\xi_{2,+}=0,
 \ee
in the formulas providing the expression of $W_{\!\text{AW}}$.
This leads to 
\be 
c_3 = 0,\quad 
\beta=q^{\frac12}, \quad \rho_8=\rho_{10} - \sum_{j=0}^7 \sum_{\ell=0}^{8-j} q^{\frac12 j +\ell -4} \rho_j, \quad \rho_9=-(q^{\frac12}+q^{-\frac12})\rho_{10} + q^{\frac12} \sum_{j=0}^7 \sum_{\ell=0}^{7-j} q^{\frac12 j +\ell -4} \rho_j, \ee
and allows to find the operator 
\be 
 W_{\!\text{AW}}^{(1)} = A_{1}^{(1)}(z) \cT^{+} + A_1^{(1)}(z^{-1}) \cT^{-} + A_0^{(1)}(z) \cI,
\ee
that satisfies \re{type2:cond}. Two parametrizations can be used (as in \cite{tsujimoto2023rational}). In the first one has

\begin{align}
    &A_1^{(1)}(z) = \dfrac{(qz - \alpha) Q_{8}(z)}{ z^2 (1-\alpha z)(1-z^2)(1-q z^2) }, \\
    &A_0^{(1)}(z)= c_0^{(1)} +c_1^{(1)}\left(z+\frac{1}{z}\right) + c_2^{(1)}\left(z^2+\frac{1}{z^2}\right) \nonumber\\
& \phantom{A_0^{(1)}(z) =} -\frac{c_4^{(1)}}{(1-q^{-\frac12}z)(1-q^{-\frac12}z^{-1})} +\frac{c_5^{(1)}}{(1+q^{-\frac12}z)(1+q^{-\frac12}z^{-1})},
\end{align}
and
\be
Q_{8}(z) = \sum_{j=0}^{8}\eta_j z^j
\ee
with $\eta_{8}=\alpha^2 q^3 \eta_0$ and
\ba
&
c_1^{(1)}=
\alpha q \eta_1 + (\alpha q)^{-1}\eta_7, \quad 
c_2^{(1)} = q  \left( 1+q\right)\alpha\eta_0,\\
&\displaystyle c_4^{(1)} = \dfrac{1}{2}q^{\frac32}\sum_{j=0}^8 q^{-\frac12 j}\eta_j, \quad 
 c_5^{(1)} = \dfrac{1}{2} q^{\frac32}\sum_{j=0}^8 (-q^{\frac12})^{-j}\eta_j.
\ea

We now introduce another parametrization $\varepsilon_1,\varepsilon_2,\ldots,\varepsilon_8$ by
\begin{equation}
\eta_k = (-1)^k q^{k/2}\sigma_k\,\eta_0
\qquad (k=1,2,\ldots,8),
\label{epsilon_para}
\end{equation}
where $\sigma_k=\sigma_k(\varepsilon_1,\varepsilon_2,\ldots,\varepsilon_8)$ denotes the $k$-th elementary symmetric polynomial in $\varepsilon_1,\varepsilon_2,\ldots,\varepsilon_8$, defined by
\[
\sigma_k(\varepsilon_1,\ldots,\varepsilon_8)
=
\sum_{1\le i_1<\cdots<i_k\le 8}
\varepsilon_{i_1}\cdots \varepsilon_{i_k}.
\]

We thus have
\be
\alpha\beta = q\, \varepsilon^{\frac12}_{1}\varepsilon^{\frac12}_{2}\varepsilon^{\frac12}_{3}\varepsilon^{\frac12}_{4}\varepsilon^{\frac12}_{5}\varepsilon^{\frac12}_{6}\varepsilon^{\frac12}_{7}\varepsilon^{\frac12}_{8} \label{cond:alpha_beta}\ee
from the condition $\eta_8=(q\alpha\beta)^2 \eta_0 = q^4 \sigma_8\eta_0$.
In this parametrization (\ref{epsilon_para}),
one has instead:
\begin{align}
&A_1^{(1)} (z) =
\dfrac{1}{q z^2}\dfrac{\prod_{j=1}^8 ( \varepsilon^{\frac12}_j-q^{\frac12} z)}{(1-q^{\frac12} z \prod_{j=1}^8 \varepsilon^{\frac12}_j)}\dfrac{\prod_{j=1}^8(1-q^{\frac12}\,\varepsilon_j z)}{(1-z^2)(1-q z^2)} ,\\
    &A_0^{(1)}(z)= c_0^{(1)} + c_1^{(1)}\left(z+\frac{1}{z}\right) + c_2^{(1)}\left(z^2+\frac{1}{z^2}\right) \nonumber\\
%+\frac{c_4^{(2)} p^2}{(p-z)(p-z^{-1})} +\frac{c_5^{(2)} p^2}{(p+z)(p+z^{-1})} 
& \phantom{A_0^{(1)}(z) =} -\frac{c_4^{(1)}}{(1-q^{-\frac12}z)(1-q^{-\frac12}z^{-1})} +\frac{c_5^{(1)}}{(1+q^{-\frac12}z)(1+q^{-\frac12}z^{-1})},
\end{align}
\ba
&& c_1^{(1)} = q^{\frac12} \prod_{\ell=1}^{8} \varepsilon^{\frac12}_{\ell} \cdot \sum_{j=1}^{k} \left(\varepsilon_j+\varepsilon_j^{-1}\right), \quad c_2^{(1)} = - (1+q) \prod_{\ell=1}^{8}\varepsilon^{\frac12}_{\ell},\\
&& 
c_4^{(1)} = \dfrac{1}{2}\prod_{j=1}^{8}(1-\varepsilon_j), \quad
c_5^{(1)} = \dfrac{1}{2}\prod_{j=1}^{8}(1+\varepsilon_j),
\ea
with $c_{0}^{(1)}$ an arbitrary parameter.

\section{The Heun operator $W^{(2)}_{\!\text{AW}}$ associated to two series of poles on the Askey-Wilson grid}
\subsection{Two additional conditions}
The operator $W_{\!\text{AW}}^{(2)}$ satisfying \re{type3:cond} is obtained from $W_{\!\text{AW}}$ by the addition of  two more conditions, namely:
\ba
\lab{W2:x'cond1}
 && W_{\!\text{AW}}^{(2)}\left\{\dfrac{1}{x-y_0}\right\}=\dfrac{\xi_{-1,+}}{x-x_0}+\dfrac{\xi_{-1,0}}{x-y_0}+\dfrac{\xi_{-1,1}}{x-y_1},\\
\lab{W2:x'cond2}
 && W_{\!\text{AW}}^{(2)}\left\{\dfrac{1}{x-y_1}\right\}=\dfrac{\xi_{-2,+}}{x-x_0}+\dfrac{\xi_{-2,0}}{x-y_0}+\dfrac{\xi_{-2,1}}{x-y_1}+\dfrac{\xi_{-2,2}}{x-y_2}.
\ea
Comparing with the computed action of $W_{\!\text{AW}}$ on the elementary rational functions $\frac{1}{x-y_0}$ and $\frac{1}{x-y_1}$ given in \eqref{W:action_y0} and \eqref{W:action_y1}, we see that we must impose 
\be \xi_{-1,-1}=\xi_{-1,0}'=\xi_{-2,0}'=0 \ee
in \re{W:action_y0} and \re{W:action_y1}
which leads to
\begin{align}
   c_3=0, \qquad  \sum_{j=0}^{10} \beta^j \rho_j =0, \qquad \sum_{j=0}^{10} q^{10-j} \beta^j \rho_j =0.
   \label{type2:cond-eta}
\end{align}
\subsection{Construction of $  W_{\!\text{\mdseries AW}}^{(2)}$}
Inserting the relations \eqref{type2:cond-eta} in the expression of $W_{\!\text{AW}}$, we obtain the operator 
 
\be
  W_{\!\text{AW}}^{(2)} = A_{1}^{(2)}(z) \cT^{+} + A_1^{(2)}(z^{-1}) \cT^{-} + A_0^{(2)}(z) \cI,\label{w2}
\ee
where
\ba
&& A_1^{(2)}(z) = \dfrac{(qz - \alpha) (qz - \beta)  Q_{8}(z)}{q^2 z^2 (\alpha z - 1)(\beta z - 1)(1-z^2)(1-q z^2) \eta_0 }, \\
&& A_0^{(2)}(z) = c_0^{(2)} +c_1^{(2)}\left(z+\frac{1}{z}\right) + c_2^{(2)}\left(z^2+\frac{1}{z^2}\right) \nonumber\\
%+\frac{c_4^{(2)} p^2}{(p-z)(p-z^{-1})} +\frac{c_5^{(2)} p^2}{(p+z)(p+z^{-1})} 
&& \phantom{A_0^{(2)}(z) =} +\frac{c_4^{(2)}}{(1-q^{-\frac12} z)(1-q^{-\frac12}/z)} +\frac{c_5^{(2)}}{(1+q^{-\frac12} z)(1+q^{-\frac12}/z)},
\ea
and $Q_{8}(z)$ is a polynomial of degree 8 in $z$ defined by
\be
Q_{8}(z) = \sum_{j=0}^{8}\eta_j z^j
\ee
with $\eta_{8}=q^2 \alpha^2 \beta^2 \eta_0$, and
\ba
&c_1^{(2)}=-\alpha\beta (q \eta_0)^{-1}\eta_1 - (\alpha \beta q^2 \eta_0)^{-1}\eta_7, \quad 
c_2^{(2)} = -\alpha \beta \left( 1+q^{-1}\right),\\
&\displaystyle c_4^{(2)} = \dfrac{1}{2 \eta_0}\sum_{j=0}^8 q^{-\frac{j}{2}}\eta_j, \quad 
 c_5^{(2)} = \dfrac{1}{2\eta_0} \sum_{j=0}^8 (-q)^{-\frac{j}{2}}\eta_j.
\ea

\subsection{Sufficiency}

The operator $W_{\!\text{AW}}^{(2)}$ was obtained by solving the necessary conditions imposed by the prescribed actions on the five elementary rational functions
\begin{equation}
    \frac{1}{x-x_0},\quad\frac{1}{x-x_1},\quad\frac{1}{x-x_2},\quad\frac{1}{x-y_0},\quad\frac{1}{x-y_1},
\end{equation}
with the poles on Askey-Wilson grids as per \eqref{xnyn}.
This completely fixed the functions $A_{1}^{(2)}(z)$ and $A_0^{(2)}(z)$ entering in the defining form \eqref{w2} of $W_{\!\text{AW}}^{(2)}$. That these conditions are also sufficient is then shown by the fact that the actions on the generic elementary rational functions of the two pole sequences can be explicitly computed and read: 
\begin{align}
&    W_{\!\text{AW}}^{(2)}\left\{\dfrac{1}{x-x_{n}}\right\} = \dfrac{\hat\xi_{n,-1}}{x-y_{0}}+\dfrac{\hat\xi_{n,0}}{x-x_{0}}+\dfrac{\hat\xi_{n,1}}{x-x_{n-1}}+\dfrac{\hat\xi_{n,2}}{x-x_{n}}+\dfrac{\hat\xi_{n,3}}{x-x_{n+1}},\\
&    W_{\!\text{AW}}^{(2)}\left\{\dfrac{1}{x-y_{n}}\right\} = \dfrac{\hat\xi_{n,-1}'}{x-x_{0}}+\dfrac{\hat\xi_{n,0}'}{x-y_{0}}+\dfrac{\hat\xi_{n,1}'}{x-y_{n-1}}+\dfrac{\hat\xi_{n,2}'}{x-y_{n}}+\dfrac{\hat\xi_{n,3}'}{x-y_{n+1}},
\end{align}
where
\begin{align}
&    \hat\xi_{n,-1}=\dfrac{-(1-\alpha\beta/q)\sum_{j=0}^8 \eta_j \beta^{4-j}}{(1-q^{n-1}\alpha\beta )(1-q^{-n-1}\beta/\alpha )(1-\alpha/\beta)q^2 \eta_0}, \\
&  \hat\xi_{n,0}=\dfrac{-(1-\alpha\beta/q)\sum_{j=0}^8 \eta_j \alpha^{4-j}}{(1-q^{n-1}\alpha^2)(1-q^{-n-1})(1-\beta/\alpha)q^2 \eta_0} ,\\
& \hat\xi_{n,1}=\dfrac{(1-q^{-n})(1-q^{-n}\beta/\alpha)\sum_{j=0}^8 \eta_j q^{(j-4)(n-1)}\alpha^{j-4}}{(1-q^{n-1}\alpha^2)(1-q^{n-1} \alpha\beta) (1-\alpha^{-2}q^{-2n})(1-\alpha^{-2}q^{-2n+1}) \, q^2 \eta_0} , \\
& \hat\xi_{n,2}= -\left(q^{\frac12}+q^{-\frac12}\right)\left( \alpha^2 q^{2n} + \frac{1}{\alpha^2 q^{2n}}\right)\frac{\alpha\beta}{q^{\frac12}}
- \left( \alpha q^n+\frac{1}{\alpha q^n}\right)\left( \alpha \beta q^{\frac12} \eta_1 + \frac{\eta_7}{\alpha\beta q^{\frac12}}\right)\frac{1}{q^{\frac32}\eta_0}
\nonumber \\
&\qquad
-\dfrac{\displaystyle
\left(q^{\frac12}+q^{-\frac12}\right)\sum_{j=0}^{4} q^{2-j}\eta_{2j} + \left(\alpha q^n+\dfrac{1}{\alpha q^{n}}\right)\sum_{j=0}^{3} q^{\frac32-j}\eta_{2j+1}
 }{\alpha^{-2}q^{-2n}(1-\alpha^2 q^{2n-1})(1-\alpha^2 q^{2n+1}) q^{\frac32} \eta_0} ,\\
& \hat\xi_{n,3}= \dfrac{(1-q^{n}\alpha^2)(1-q^{n}\alpha\beta)\sum_{j=0}^8 \eta_j q^{(4-j)(n+1)}\alpha^{4-j}}{(1-q^{-n-1})(1-q^{-n-1} \beta/\alpha) (1-\alpha^{2}q^{2n})(1-\alpha^{2}q^{2n+1}) \, q^2 \eta_0} ,
\end{align}
and $\hat\xi_{n,\ell}'=\hat\xi_{n,\ell} \big|_{\alpha \leftrightarrow \beta}$ for $\ell \in \{-1,0,1,2,3\}$. Note that when $n=0$, both $\hat\xi_{n,1}$ and $\hat\xi_{n,1}'$ are zero.

This completes the proof that the operator $W^{(2)}_{AW}$ constructed in this section  satisfy \eqref{type3:cond}.

\begin{remark}
    Comparing with the expression of $W^{(1)}_{\!\text{AW}}$, we note that one of the signs in front of the four terms in \( A_0^{(1)} \) is negative, which distinguishes it from \( A_0^{(2)} \).
\end{remark}

\begin{remark}
  A point could be made here about the situation (a priori excluded) where the two series of poles are specialized into a single one, i.e., $\alpha=\beta$. One could then have
\begin{align}
  & W_{\!\text{AW}}^{(2)}\left\{\dfrac{1}{x-x_n}\right\} \biggr|_{\beta \to \alpha}= 
\dfrac{\xi_{-1}''}{(x-x_0)^2}+\dfrac{\xi_{0}''}{x-x_0}+\dfrac{\xi_{1}''}{x-x_{n-1}}+\dfrac{\xi_{2}''}{x-x_{n}}+\dfrac{\xi_{3}''}{x-x_{n+1}}.
\end{align}
Consequently an undesirable pole of second order would appear. In such an instance, a further parameter specialization would  be needed to make this term vanish and this would lead to a specialized $W_{\!\text{AW}}^{(1)}$ operator.
\end{remark}

\section{Identification with the Takemura operator $\cA^{\langle 1\rangle}$}

We will now connect the Heun operators $W_{\!\text{AW}}^{(1)}$ and $W_{\!\text{AW}}^{(2)}$ to the Takemura Hamiltonian. We shall start by considering the latter.

\subsection{$W_{\!\text{\mdseries AW}}^{(2)}$ and $\cA^{\langle 1\rangle}$}

Let us introduce the gauge transformation:
\be \hat W_{\!\text{AW}}^{(2)} = \psi(z)^{-1} W_{\!\text{AW}}^{(2)} \psi(z) 
\ee
where $\psi(qz)=q^2 z^2 \dfrac{(\alpha z - 1)(\beta z - 1)}{(qz-\alpha)(qz-\beta)}\psi(z)$ holds.
We then obtain
\be
\hat W_{\!\text{AW}}^{(2)} = \hat A_1^{(2)}(z) \cT^{+}+\hat A_1^{(2)}(z^{-1}) \cT^{-} + \hat A_0^{(2)}(z)\cI
\ee
where $\hat A_0^{(2)}(z) = A_0^{(2)}(z)$ and
\be
\hat A_1^{(2)}(z) = \dfrac{\psi(qz)}{\psi(z)} A_1^{(2)}(z)=\dfrac{\sum_{j=0}^{8}\eta_j z^j}{(1-z^2)(1-qz^2)\eta_0},
\ee
with $\eta_8 = q^2\alpha^2\beta^2\eta_0$.

In the alternate parametrization (\ref{epsilon_para}),
the functions $\hat A_1^{(2)} (z)$ and $\hat A_0^{(2)}(z)$ that define $\hat W_{\!\text{AW}}^{(2)}$ read: 
\be \hat A_1^{(2)} (z) =
\dfrac{\prod_{j=1}^8(1-q^{\frac12}\,\varepsilon_j z)}{(1-z^2)(1-q z^2)},\ee
\begin{align}
\hat A_0^{(2)} (z) =& c_0^{(2)} + c_1^{(2)}\left(z+\frac{1}{z}\right) + c_2^{(2)}\left(z^2+\frac{1}{z^2}\right) \nonumber\\
&+\frac{c_4^{(2)} }{(1-q^{-\frac12}z)(1-q^{-\frac12}z^{-1})} +\frac{c_5^{(2)} }{(1+q^{-\frac12}z)(1+q^{-\frac12}z^{-1})},
\end{align}
where
\ba
&& c_1^{(2)} = q^{\frac12} \prod_{\ell=1}^{8} \varepsilon^{\frac12}_{\ell} \cdot \sum_{j=1}^{k} \left(\varepsilon_j+\varepsilon_j^{-1}\right), \quad c_2^{(2)} = - (1+q) \prod_{\ell=1}^{8}\varepsilon^{\frac12}_{\ell},\\
&& 
c_4^{(2)} = \dfrac{1}{2}\prod_{j=1}^{8}(1-\varepsilon_j), \quad
c_5^{(2)} = \dfrac{1}{2}\prod_{j=1}^{8}(1+\varepsilon_j),
\ea
and $c_{0}^{(2)}$ is an arbitrary parameter.
\textit{This coincides with Takemura's $\cA^{\langle 1\rangle}$ operator in our notation}.

\subsection{$W_{\!\text{\mdseries AW}}^{(1)}$ and $\cA^{\langle 1\rangle}$}

Through reparametrization and an appropriate gauge transformation, two operators $W_{\!\text{AW}}^{(2)}$ and $W_{\!\text{AW}}^{(1)}$ can be transformed into each other. For instance, the following relation holds between $W_{\!\text{AW}}^{(1)}$ and $W_{\!\text{AW}}^{(2)}$:
\begin{align}
    W_{\!\text{AW}}^{(2)}\bigg|_{\varepsilon_8\to \varepsilon^{-1}_8}
    = \varepsilon_8^{-1} \widetilde\psi(z) W_{\!\text{AW}}^{(1)} \widetilde\psi(z)^{-1}, 
    \label{relation:W2_W1}
\end{align}
% (z; {\color{blue}\varepsilon^{\frac12}}_1, {\color{blue}\varepsilon^{\frac12}}_2, {\color{blue}\varepsilon^{\frac12}}_3, {\color{blue}\varepsilon^{\frac12}}_4, {\color{blue}\varepsilon^{\frac12}}_5, {\color{blue}\varepsilon^{\frac12}}_6, {\color{blue}\varepsilon^{\frac12}}_7, {\color{blue}\varepsilon^{\frac12}}_8) 
with the function satisfying
\begin{align}
    \widetilde \psi(qz)=\dfrac{(1-\alpha q^{-1}z^{-1})(1-\beta q^{-1}z^{-1})(1-\alpha\beta \varepsilon_8^{-1}q^{-\frac12}  z )(1-\varepsilon_8 q^{\frac12} z)}{(1-\alpha z)(1-\beta z)(1-\alpha\beta \varepsilon_8^{-1} q^{-\frac32} z^{-1} )(1- \varepsilon_8 q^{-\frac12}z^{-1})}\widetilde\psi(z).
\end{align}

It follows from this observation that the Heun operator $W_{\!\text{AW}}^{(1)}$ based on a single series of pole also leads to the Takemura Hamiltonian $\cA^{\langle 1\rangle}$

\section{Concluding remarks}
The one-particle Hamiltonian of the Ruijsenaars–van Diejen model originates from a parameter space endowed with an $E_8$-type symmetry\cite{ruijsenaars2015relHeun}. This symmetry appears prominently in the broader landscape of elliptic difference equations, including the geometric symmetry underlying the elliptic discrete Painlevé equation. Here we have been concerned with the $q$-difference degeneration of this model. The Hamiltonian $\mathcal{A}^{(1)}$ identified by Takemura in this limit has been the central object of attention of this paper, the goal being to show that this operator $\mathcal{A}^{(1)}$ is in some sense a $q$-Heun operator. 

This could indeed be confirmed by exploiting the general idea that Heun operators can be defined as operators endowed with a raising property when acting on sets of basic functions attached to appropriate grids.

In summary, the Heun feature of the Hamiltonian $\mathcal{A}^{(1)}$ can be underscored as follows. One first identifies the most general second order $q$-difference operator $W_{AW}^{(2)}$ that acts as a raising operator on two sets of elementary rational functions with poles on two different Askey-Wilson grids. $\mathcal{A}^{(1)}$ is then found from this $q$-difference operator upon performing a gauge transformation and some reparametrization. The gauge transformation eliminates in fact a parameter from the second series of elementary rational functions and affects, in fine, the ``raising rule'' for $\mathcal{A}^{(1)}$ albeit without changing its Heun operator nature. Alternatively, one may proceed to obtain $\mathcal{A}^{(1)}$ by characterizing the most general second order $q$-difference operator $W_{AW}^{(1)}$ that acts as a raising operator on one set of elementary rational functions with poles on an Askey-Wilson grid since it was further shown that $W_{AW}^{(1)}$ can be obtained from $W_{AW}^{(2)}$ again by a gauge transformation and reparametrization.

Takemura has also identified a second Hamiltonian $\mathcal{A}^{(2)}$. Since this operator is obtained from $\mathcal{A}^{(1)}$ by a degeneration it will also be a Heun operator. This degeneration amounts to considering the poles on the $q$-linear grid instead of the Askey-Wilson one. The independent characterization of the rational Heun operator defined on that grid is important and proceeds in its own ways. It will be carried out in a future publication.

The present study clearly raises two major questions. The first asks whether there exists or not an equivalent definition of the rational Heun operators that would align with the notion of algebraic Heun operators that applies to the Heun operators that are in correspondence with hypergeometric orthogonal polynomials. Recent advances related to the introduction of Leonard trios and their connections to rational functions offer promising perspectives in that respect \cite{CGLMVW_Trios}. The second, also quite challenging, is whether the Hamiltonians for $N \geq 2$ admit a definition as Heun operators through raising conditions. We hope to pursue these matters in the future.

\section*{Acknowledgments}
This work has been sponsored by a Québec-Kyoto cooperation grant from the Ministère des Relations Internationales et de la Francophonie of the Quebec Government.
The research of ST is supported by JSPS KAKENHI (Grant Number 24K00528). LV is funded in part through a discovery grant of the Natural Sciences and Engineering Research Council (NSERC) of Canada. 
 AZ is now supported by the Ministry of Science and Higher Education of the Russian Federation (agreement no. 075–15–2025–343).

\section*{Conflict of interest}
On behalf of all authors, the corresponding author states that there is no conflict of interest.

\section*{Data availability}
This manuscript has no associated data.

\appendix

 \section{Derivation of $\hat W_{\!\text{AW}}^{(2)}$ and $\hat\cW$ from the Ruijsenaars-van Diejen Operator}

The Ruijsenaars-van Diejen operator is presented in \cite{atai_noumi2023} as
\begin{align*}
  {\cR} = A^{+}(z) \cT +A^{-}(z) \cT^{-1} + A^{0}(z)\cI
\end{align*}
where
\begin{align*}
 &A^{+}(z) = \dfrac{\prod_{0\leq s \leq 7} \theta(a_s z;p)}{\theta(z^2;p)\theta(qz^2;p)}, 
\qquad A^{-}(z) = A^{+}(z^{-1}), \\
 & A^{0}(z)=\dfrac{1}{2}\left(A_0^{0}(z) +A_1^{0}(z) +A_2^{0}(z) +A_3^{0}(z)\right), \\
 & A_j^{0}(z) = \dfrac{L_j}{\theta(t;p) \theta(q^{-1}t;p)} \dfrac{\theta(c_j q^{-\frac12}t z ;p)\theta(c_j q^{-\frac12}tz^{-1} ;p)}{\theta(c_j  q^{-\frac12} z ;p)\theta(c_j  q^{-\frac12} z^{-1} ;p)}\prod_{0 \leq s \leq 7} \theta(c_j  q^{-\frac12} a_s;p),\\
 &\theta(z;q) = (q/z;q)_{\infty}(z;q)_{\infty}, 
\end{align*}
and 
$L_0=L_1 =1, L_2 = t q^{-2} p^{-1} \prod_{0 \leq s \leq 7}a_s, L_3 = p^2 L_3^{-1}$, $c_0=1,c_1=-1, c_2=p^{-\frac12}, c_3 = -c_2^{-1}$.

The operator $\hat W_{\!\text{AW}}^{(2)}$, a \(q\)-analogue obtained from the raising property condition for the two-series of poles, can be derived by taking the \(p \to 0\) limit of the Ruijsenaars–van Diejen operator \(\mathcal{R}\), as shown below,
\begin{align*}
& \lim_{p\to 0} A^{+}(z) = \hat A^{(2)}(z), \\
& \lim_{p\to 0} A^{-}(z) = \hat A^{(2)}(z^{-1}), \\
&\lim_{p\to 0} \left\{ A^{0}(z)  - \dfrac{ t \prod_{0 \leq s \leq 7} a_s^{\frac12}}{(1-t)(1-q^{-1}t)q^2} \dfrac{1}{p} \right\} = \hat A_0^{(2)}(z),
\end{align*}
with the replacement of the parameters $a_j \to q^{\frac12}\, \varepsilon_{j+1}$ and some constant $\hat c_0$.

\par

As discussed in our previous paper \cite{tsujimoto2023rational}, the classical rational Heun operator \(\hat \cW=\hat B(z) (\cT_q -\cI)+\hat B(z^{-1}) (\cT_q^{-1} -\cI) + \hat \gamma_{00}\cI\) can be derived from \(\hat W_{\!\text{AW}}^{(1)}\). Accordingly, by applying the relation (\ref{relation:W2_W1}) to transform \(\hat W_{\!\text{AW}}^{(2)}\), which is obtained via the limiting procedure described above, into \(\hat W_{\!\text{AW}}^{(1)}\), we obtain a derivation of \(\hat{\cW}\) from the Ruijsenaars-van Diejen operator.  
Moreover, an alternative derivation procedure can be considered: it is also possible to derive \(\hat{\cW}\) directly from the Ruijsenaars-van Diejen operator \(\mathcal{R}\) by choosing the parameters as 
\begin{align*}
 a_0 a_1 a_2 a_3 a_4 a_5 a_6 a_7= q p, \quad     
 a_0 a_1 a_2 a_3 a_4 a_5 a_6^2= q^{\frac32},    
%  a_6 a_7^{-1} = q^{\frac12} p^{-1},  
\end{align*}
and taking the limit as \( p \to 0 \):
\begin{align*}
& \lim_{p\to 0} q^{-1}p\,A^{+}(z) = \hat B(z), \\
& \lim_{p\to 0} q^{-1}p\, A^{-}(z) = \hat B(z^{-1}), \\
&\lim_{p\to 0} \left\{q^{-1}p \,A^{0}(z) -\frac{t a_0^2 a_1^2 a_2^2 a_3^2 a_4^2}{q^4 (q - t) (1 - t)} \dfrac{1}{p} \right\} = -\hat B(z)-\hat B(z^{-1})+\hat \gamma,
\end{align*}
with the replacement of the parameters $a_j \to  q^{\frac12}\, \varepsilon_{j+1}$ and some constant $\hat \gamma$.

\bibliographystyle{unsrt} 
%\bibliographystyle{unsrtinur} 
% formatting style : order of apparition numbering, number citations, 
% author initials, arXiv clickable URLs
\bibliography{ref_takham1.bib} 

\end{document}